**Journal of Financial Risk Management, 2025, 14(3), 304-324**
https://www.scirp.org/journal/jfrm
ISSN Online: 2167-9541
ISSN Print: 2167-9533

# Enhancing Efficiency of Pension Schemes through Effective Risk Governance: A Kenyan Perspective

## Sylvester Willys Namagwa

Department of Finance and Accounting, Faculty of Business and Management Science, University of Nairobi, Nairobi, Kenya
Email: sylvesternamagwa@gmail.com

**How to cite this paper:** Namagwa, S. W. (2025). Enhancing Efficiency of Pension Schemes through Effective Risk Governance: A Kenyan Perspective. *Journal of Financial Risk Management, 14,* 304-324.
https://doi.org/10.4236/jfrm.2025.143017

**Received:** August 20, 2025
**Accepted:** September 20, 2025
**Published:** September 23, 2025

## Abstract

The efficiency of Kenya's pension schemes invites elevated interest owing to the increasing pension contribution amounts and the expectation that benefits paid out of these schemes would protect members from old age poverty. The study investigates the intervening effect of risk management on the relationship between corporate governance and the efficiency of pension schemes in Kenya. The study employs panel data consisting of 896 observations from 128 schemes in a sample period from 2015 to 2021. The study finds that risk management significantly mediates the relationship between employee representatives on the board of trustees, as a component of corporate governance, and the efficiency of pension schemes. Consequently, the mediation effect of risk management indicates that when employee representatives are involved in governance, the presence of strong risk management practices ensures that their contributions lead to improved efficiency. Risk management, therefore, serves as a critical safeguard that enables governance structures to function more effectively and contribute to the overall performance of the scheme.

## 1. Introduction

### 1.1. Background

Globally, retirement assets are growing remarkably, suggesting a renewed focus on retirement security. While the OECD countries registered an 8.5% growth in their retirement assets in 2024 (OECD, 2024), Kenya's assets hit 20% (KNBS, 2025). However, pension schemes invariably face a dilemma on where and how to manage their savings to sustainably achieve the highest level of efficiency in a risk-return trade-off (Berardi & Tebaldi, 2024). They strive for a balance between growth-oriented investments and risk management strategies like overfunding to ensure both long-term growth and security for retirees (Barone-Adesi et al., 2025). Effective risk management is crucial for ensuring the long-term sustainability of schemes (Bocchialini et al., 2025) and the well-being of their members to bridge protection gaps in retirement benefits (OECD, 2025).

Risk management involves the initiatives that firms take to price, influence, and accurately reward their exposure to a risky business environment. Jefferson (2023) argues that effective risk management in pension schemes obtains if participants wield financial expertise. However, it is not practicable for every scheme member to have financial literacy (Castagno et al., 2025), especially in developing countries like Kenya. The emergence of corporate governance as the bit and bridle that investors, through their boards, use to rein in on their managers (OECD, 2023) suffices for scheme members who have no financial literacy. Boards can implement strong governance practices that leverage innovative strategies like overfunding (Barone-Adesi et al., 2025) to improve retirement outcomes while safeguarding the interests of their members.

Further, boards make decisions that structure risk profiles of their firms to influence efficiency and reward inputs (Cheng, 2024). Corporate governance has a risk management toolkit that helps boards to weave an optimal trade-off between the cost of risk control and the likely losses from risk crystallization (Anton et al., 2025; Bonyi & Stewart, 2019; Masanja, 2021). This study investigates the intervening role of risk management in the relationship between corporate governance and efficiency of pension schemes in Kenya, and demonstrates that risk management is not just a compliance exercise, but a vital component of good governance that directly contributes to the success and sustainability of pension schemes.

### 1.2. Problem Statement

Risk management in the Kenyan pension industry is yet to attract profound academic interest despite facing a significant demographic shift with both an aging population and increasing urbanization (World Bank Group, 2019) pointing to a looming need for sufficient retirement income. Efficient pension schemes are crucial for ensuring sufficient income replacement for retiring scheme members (Berardi & Tebaldi, 2024), and therefore the failure of these schemes portends a humiliating retirement life for the elderly, most of whom rely on retirement income as their only source of income (Ouma et al., 2025). Remarkably, failure of governance systems to manage risks is increasingly linked to corporate failures (Bonyi & Stewart, 2019; Khan et al., 2018; Masanja, 2021; OECD, 2014).

Pension schemes are involved in long-term investments with the possibility of asset values varying from their projections, making risk management vital for achieving actuarial fairness in benefit pay-outs (Jefferson, 2023). Further, when Boards of Trustees effectively manage risks, the returns to the scheme are optimized (Katto & Musaali, 2019). However, just like Jefferson (2023), Kiwanuka (2019) observes that the effectiveness of the trustees depends majorly on their skillsets and the regulation regime under which they operate. How, then, should the boards be structured to be effective? The traditional corporation runs on a contract between its owners and their managers, against which its governance structures are framed. Yet the business of pension schemes in Kenya confounds this bipartite governance template, and instead introduces an intricate network of relationships across the value web of the schemes' business operations with a retinue of service providers and advisors with multiple decision nodes that attract risk exposure.

Studies conducted to demystify the role of risk management in the efficiency of firms arise with various gaps and with inconclusive results as of the role of risk management in the relationship between corporate governance and the efficiency of pension schemes in Kenya. For instance, some find a negative relationship (González et al., 2020; Naibaho & Mayayogini, 2023) while others find positive links (Al-Nimer et al., 2021; Florio & Leoni, 2017; Malik et al., 2020; Ondiba et al., 2024; Yang et al., 2018). The reminder of the paper is structured as follows: The next section will cover an empirical review followed by research methodolgy, findings and discussion, conclusions and recommendations, and end with areas of further research.

## 2. Empirical Review

Al-Nimer et al. (2021) examine the relationship between risk management and firm performance on 228 financial firms in Jordan and find a positive relationship. However, this study presents three gaps. First, the reliance upon cross-sectional data whose character does not provide room to correct for unobserved endogeneity. Second, the study ignores the conceptual utility of regulation in financial firms despite its role in structuring and influencing their risk profiles. Third, Jordan's socio-economic parameters are different from Kenya's, including the size of the economy and the sophistication of the financial sectors. Florio and Leoni (2017) find a positive correlation between ERM implementation and firm performance, both in terms of financial performance and market evaluation. However, the study's findings are specific to Italian listed companies, and the results might not be generalizable to other contexts, such as pension schemes.

González et al. (2020) find that ERM has no relation with performance of listed firms in Spain. Yet the study shows a conceptual limitation with regard to missing variables that could potentially mediate the relationship between ERM and firm performance. Theoretically, firms create strategic value not only when they are sensitive to the operating environment, but also when they optimize their resources (Anton et al., 2025). In contrast, Yang et al. (2018) find a positive relationship between risk management practices and performance of Small and Medium Enterprises (SMEs) in Pakistan. However, the singular focus on SMEs predisposes the study to sampling bias against firms of different dispositions. Ondiba et al. (2024) find that risk management has a significant positive influence on performance of agricultural State corporations in Kenya breeding the contextual contrast with the pension industry.

Malik et al. (2020) find a significant and positive influence of ERM on firm performance particularly through board-level risk committee as a governance mechanism on listed firms in the UK. However, the study exhibits a gap regarding the contextual disparity between UK and Kenya, the former being a developed country with a more sophisticated governance system. Naibaho and Mayayogini (2023) find a mixed result while determining the impact of various risk management parameters on firm performance with corporate governance as the moderating variable. This study had limitations with regard to the inclusion of firms which did not meet the reasearch criteria in addition to using foreign proxies as a prdictor for corporate governance.

Mitei et al. (2025) determine the effect of management innovativeness and growth of pension schemes in Nairobi and find that management innovativeness has a significant positive influence on the growth of pension schemes. The study emerges with a conceptual gap in neglecting the role of risk management, the assessment of which determines the adoption of innovations in firms (Anton et al., 2025). The current study bridges the aforementioned gaps by examining the mediating role of risk management in the relationship between corporate governance and efficiency of pension schemes in Kenya.

## Conceptual Framework

The conceptual model, depicted in Figure 1, illustrates the relationships between the variables within the research framework. Corporate governance as the dependent variable is operationalised based on board diversity i.e. proportion of female members, scheme members, top management, and independent members on the board of trustees.

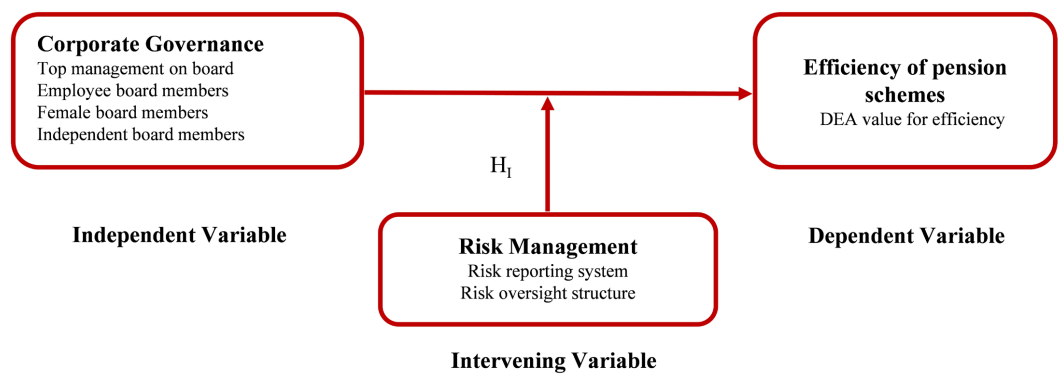

**Figure 1.** Conceptual model.

Risk management, as the intervening variable, structures the schemes' risk profiles through risk oversight structure and risk reporting system and it is measured upon the degree of development of these infrastructures in each scheme. Risk management in each scheme was quantified on a scale of 1 - 5 based on disclosures about the presence of risk management oversight structure and culture, control systems, reporting and communication systems, risk strategy, and risk assessment initiatives. Efficiency as the dependent variable is determined through Data Envelopment Analysis (DEA) using annual administration and investment costs as inputs, while return on investments and change in scheme assets as outputs.

## 3. Research Methodology

Quantitative research design was adopted to describe the position of variables and establish the extent of their relationships while panel research design is adopted to track performance changes among pension schemes from 2015 to 2021. The study used stratified sampling along Occupational schemes, Individual schemes, and Umbrella schemes to sample 128 pension schemes from the 1189 schemes in the Retirement Benefits Authority's register, all providing 896 observations. The study used secondary data from various reports and audited financial statements of the sampled schemes.

Multiple regression models were used to assess the intervening effect of risk management in the relationship between corporate governance and the efficiency of pension schemes, adopting Baron and Kenny (Baron & Kenny, 1986). Here, four steps were pursued to examine the intervening effect of risk management using the models below.

The first step seeks to evaluate the independent and dependent variables' correlations to determine if there exists a relationship that can be mediated.

$$EF_i = \beta_0 + \beta_i CG + \varepsilon_i \tag{3.1}$$

$$RRAI = \beta_0 + \beta_i CG + \varepsilon_i \tag{3.2}$$

**where:**

$EF_i$/RRAI: Pension schemes/technical efficiency scores.

Mgt: Top management on board.

$\beta_0$: Regression constant or intercept.

Mb: Employee representatives on board.

$\beta_i$: Regression coefficients of variable $i$.

Fm: Female board members.

$\varepsilon_i$: Error term.

Im: Independent board members.

In the second step, regression analysis is done to examine the relationship between corporate governance and risk management, ignoring the dependent variable, efficiency of pension schemes using the following model:

$$Rm = \beta_0 + \beta_i CG + \varepsilon_i \tag{3.3}$$

**where:**

Rm: The level of development of risk management architecture

In the third step, regression analysis is done to establish the relationship between efficiency of pension schemes and risk management while ignoring the independent variable, corporate governance using this model:

$$EF_i = \beta_0 + \beta_1 Rm + \varepsilon_i \quad (3.4)$$

$$RRAI = \beta_0 + \beta_1 Rm + \varepsilon_i \quad (3.5)$$

where:

$EF_i$, $\beta_0$, $\beta_i$, $\varepsilon_i$ and RRAI is as defined above.

The final step in analysing the intervener involves a joint regression analysis of the three variables, efficiency of pension schemes, risk management and corporate governance.

$$EF_i = \beta_0 + \beta_1 CG + \beta_2 Rm + \varepsilon_i \quad (3.6)$$

$$RRAI = \beta_0 + \beta_1 CG + \beta_2 Rm + \varepsilon_i \quad (3.7)$$

where:

$EF_i$, $\beta_0$, $\beta_i$, $\varepsilon_i$ and RRAI are as defined above.

The study employed $R^2$ to assess the intervening variable variation due to the effects of the independent variable. To assess the model fit, the F-Test was used to test the significance of the model while T-Test was used to evaluate the significance of the beta coefficient of the predictor variable. If the relationship between Corporate Governance and Efficiency of pension schemes became statistically significant when Risk Management is controlled for; then full mediation was inferred. In case the relationship between Corporate Governance and Efficiency of pension schemes became statistically significant when Risk Management is controlled for, then partial mediation was presumed to have obtained. In other words, Intervention would obtain if corporate governance predicted efficiency of pension schemes, corporate governance predicted risk management, risk management predicted efficiency of pension schemes, and corporate governance predicted efficiency of pension schemes when risk management was in the model.

If the *p*-value from the mediation analysis shows a value less than 0.05, the null hypothesis ($H_{01}$) was rejected. This would mean that risk management has a significant intervening (mediating) effect on the relationship between corporate governance and the efficiency of pension schemes.

## 4. Findings and Discussion

This research aimed to establish the intervening effect of risk management on the association between corporate governance and efficiency of pension schemes in Kenya. This objective integrates the various aspects of corporate governance along with the roles of risk management to determine their combined influence on efficiency as shown in the hypothesis and sub-hypotheses below.

**$H_{01}$:** Risk management has no intervening effect on the relationship between corporate governance and efficiency of pension schemes in Kenya.

**$H_{01a}$:** Risk management has no intervening effect on the relationship between top management on board and efficiency of pension schemes in Kenya.

$H_{01b}$: Risk management has no intervening effect on the relationship between employee board members and efficiency of pension schemes in Kenya.

$H_{01c}$: Risk management has no intervening effect on the relationship between female board members and efficiency of pension schemes in Kenya.

$H_{01d}$: Risk management has no intervening effect on the relationship between independent board members and efficiency of pension schemes in Kenya.

Multiple regression analyses were carried out in four phases, with the significance of the coefficients assessed at each stage. The first two phases utilized simple linear regression, whereas the third and fourth steps used multiple regressions. The first step evaluates the independent and dependent variables' correlations to demonstrate that the predictor and dependent variables are related. A statistically significant relationship should exist to determine that a relationship exists that can be mediated. The second step estimates the association between the independent and mediator variables to show that the independent variable and the mediator are correlated. This stage essentially requires treating the mediator as an outcome variable.

The third step involves the control for the independent variable and estimate the connection among the intervening and the criterion variable to show that the mediator affects the dependent variable. The final step demonstrates that the connection among the independent and criterion variables is insignificant in the mediator's presence. The effect of the independent variable on the dependent variable should be zero when controlling for the mediator, indicating that the mediator mediates the independent-dependent variable relationship. The Baron and Kenny (1986) approach for testing mediation presumes that the independent variable predicts the dependent variable significantly.

$H_{01a}$: Risk management has no intervening effect on the relationship between top management on board and efficiency of pension schemes in Kenya.

The objective of the study examines the intervening effect of risk management on the relationship between top management on the board and the efficiency of pension schemes in Kenya by testing the above hypothesis.

**Step 1: Top Management on Board and Efficiency**

The first step examines the direct relationship between top management on board (LNMgtR) and efficiency without considering risk management as a mediator. Table 1 shows that the coefficient for LNMgtR is −0.01113, with a t-value of −0.07 and a $P$-value of 0.944. The R-squared (within) value is 0.0000, indicating that the model explains virtually none of the variation in efficiency. The relationship is statistically insignificant, suggesting that top management representation on the board does not directly influence the efficiency of pension schemes. However, significant direct mediation (Baron & Kenny, 1986) is no longer a strict requirement for establishing mediation since contemporary approaches assess the indirect effect directly (MacKinnon, Fairchild, & Fritz, 2007), such as through bootstrapping.

Table 1. Top management on board and efficiency.

| Fixed-effects (within) regression | Number of obs | = | 896 |
|---|---|---|---|
| Group variable: Scheme | Number of groups | = | 128 |
| R-sq: | Obs per group: | | |
| within = 0.0000 | min | = | 7 |
| between = 0.0027 | avg | = | 7 |
| overall = 0.0009 | max | = | 7 |
| | F(1, 767) | = | 0 |
| corr(u_i, Xb) = 0.0408 | Prob > F | = | 0.9441 |
| Efficiency | Coef. | Std. Err. | t | P > t | [95% Conf. | Interval] |
| LNMgtR | −0.01113 | 0.158567 | −0.07 | 0.944 | −0.32241 | 0.300146 |
| _cons | −0.9683 | 0.260286 | −3.72 | 0.000 | −1.47926 | −0.45734 |

### Step 2: Top Management on Board and Risk Management

The second step evaluates the relationship between top management on board (LNMgtR) and risk management (LNRM). Table 2 shows that the coefficient for LNMgtR is −0.15173, with a t-value of −3.75 and a P-value of 0.000. The R-squared (within) value is 0.0180, indicating that about 1.80% of the variability in risk management practices can be explained by top management representation on the board. The negative and statistically significant coefficient suggests that higher representation of top management on the board is associated with lower levels of risk management, implying a potential conflict or oversight issue in risk management practices.

Table 2. Top management and risk management.

| Fixed-effects (within) regression | Number of obs | = | 896 |
|---|---|---|---|
| Group variable: Scheme | Number of groups | = | 128 |
| R-sq: | Obs per group: | | |
| within = 0.0180 | min | = | 7 |
| between = 0.0167 | avg | = | 7 |
| overall = 0.0009 | max | = | 7 |
| | F(1, 767) | = | 14.07 |
| corr(u_i, Xb) = −0.3202 | Prob > F | = | 0.0002 |
| LNRM | Coef. | Std. Err. | t | P > t | [95% Conf. | Interval] |
| LNMgtR | −0.15173 | 0.040446 | −3.75 | 0.000 | −0.23113 | −0.07234 |
| _cons | 0.896432 | 0.066393 | 13.5 | 0.000 | 0.7661 | 1.026765 |

### Step 3: Risk Management and Efficiency

The third step investigates the relationship between the mediator (risk management) and the dependent variable (efficiency). Table 3 indicates that the coefficient for LNRM is 0.3843364, with a t-value of 2.77 and a P-value of 0.006. The R-squared (within) value is 0.0099, showing that risk management accounts for nearly 1% of the variability in efficiency. The positive and statistically significant coefficient suggests that better risk management practices are associated with higher efficiency in pension schemes.

Table 3. Risk management and efficiency.

| Fixed-effects (within) regression | | | Number of obs | | = | 896 |
|---|---|---|---|---|---|---|
| Group variable: Schemer | | | Number of groups | | = | 128 |
| R-sq: | | | Obs per group: | | | |
| within = 0.0099 | | | min | | = | 7 |
| between = 0.0299 | | | avg | | = | 7 |
| overall = 0.0034 | | | max | | = | 7 |
| | | | F(1, 767) | | = | 7.66 |
| corr(u_i, Xb) = −0.3433 | | | Prob > F | | = | 0.0058 |
| Efficiency | Coef. | Std. Err. | t | P > t | [95% Conf. | Interval] |
| LNIR | 0.3843364 | 0.138905 | 2.77 | 0.006 | 0.111658 | 0.657015 |
| _cons | −1.532897 | 0.212364 | −7.22 | 0.000 | −1.94978 | −1.11601 |

### Step 4: Top Management on Board, Risk Management and Efficiency

The final step assesses the relationship between top management on board and efficiency while controlling for risk management. Table 4 shows that the coefficient for LNRM remains positive and statistically significant (0.527188, $P = 0.000$), indicating that risk management continues to have a significant positive effect on efficiency when included in the model. However, the coefficient for LNMgtR is now positive (0.068863) but remains statistically insignificant ($P = 0.664$), suggesting that the direct influence of top management on efficiency becomes negligible when risk management is accounted for.

Table 4. Top management on board, risk management and efficiency.

| Fixed-effects (within) regression | | | Number of obs | | = | 896 |
|---|---|---|---|---|---|---|
| Group variable: Scheme | | | Number of groups | | = | 128 |
| R-sq: | | | Obs per group: | | | |
| within = 0.0181 | | | min | | = | 7 |
| between = 0.0034 | | | avg | | = | 7 |
| overall = 0.0025 | | | max | | = | 7 |
| | | | F(2, 766) | | = | 7.06 |
| corr(u_i, Xb) = −0.1574 | | | Prob > F | | = | 0.0009 |
| Efficiency | Coef. | Std. Err. | t | P > t | [95% Conf. | Interval] |
| LNRM | 0.527188 | 0.140364 | 3.76 | 0.000 | 0.251644 | 0.802731 |
| LNMgtR | 0.068863 | 0.158665 | 0.43 | 0.664 | −0.24261 | 0.380332 |
| _cons | −1.44089 | 0.287129 | −5.02 | 0.000 | −2.00454 | −0.87723 |

Based on these results, the study concludes that risk management does not have an intervening effect on the relationship between top management on board and efficiency. Therefore, the study fails to reject the null hypothesis $H_{01a}$ and confirms that risk management does not mediate the relationship between top management on board and the efficiency of pension schemes in Kenya. This finding highlights the complex role of risk management in bridging the gap between corporate governance practices and organizational efficiency.

$H_{01b}$: Risk management has no intervening effect on the relationship between employee board members and efficiency of pension schemes in Kenya.

The objective of the study also examines the intervening effect of risk management on the relationship between employee representatives on the board and the efficiency of pension schemes in Kenya by testing the above hypothesis.

### Step 1: Employee Reps on the Board and Efficiency

The first step evaluates the direct relationship between employee representatives on the board (LNMbR) and efficiency without considering risk management as a mediator. Table 5 shows that the coefficient for LNMbR is 0.948121, with a t-value of 7.77 and a $P$-value of 0.000. The R-squared (within) value is 0.0730, indicating that about 7.30% of the variation in efficiency can be explained by the proportion of employee representatives on the board. The positive and statistically significant coefficient suggests that higher employee representation on the board is associated with increased efficiency in pension schemes.

Table 5. Employee reps on the board and efficiency.

| Fixed-effects (within) regression | Number of obs | = | 896 |
|---|---|---|---|
| Group variable: Scheme | Number of groups | = | 128 |
| R-sq: | Obs per group: | | |
| within = 0.0730 | min | = | 7 |
| between = 0.0171 | avg | = | 7 |
| overall = 0.0037 | max | = | 7 |
| | F(1, 767) | = | 60.4 |
| corr(u_i, Xb) = −0.4191 | Prob > F | = | 0 |

| Efficiency | Coef. | Std. Err. | t | $P > t$ | [95% Conf. | Interval] |
|---|---|---|---|---|---|---|
| LNMbR | 0.948121 | 0.122001 | 7.77 | 0.000 | 0.708626 | 1.187617 |
| _cons | −0.14756 | 0.106563 | −1.38 | 0.167 | −0.35675 | 0.061633 |

### Step 2: Employee Reps on the Board and Risk Management

The second step examines the relationship between employee representatives on the board (LNMbR) and risk management (LNRM). Table 6 indicates that the coefficient for LNMbR is 0.092272, with a t-value of 2.84 and a $P$-value of 0.005. The R-squared (within) value is 0.0104, suggesting that approximately 1.04% of the variation in risk management practices can be explained by employee board representation. The positive and statistically significant coefficient indicates that higher representation of employee representatives on the board is associated with better risk management practices.

Table 6. Employee reps on the board and risk management.

| Fixed-effects (within) regression | | Number of obs | = | 896 |
|---|---|---|---|---|
| Group variable: Scheme | | Number of groups | = | 128 |
| R-sq: | | Obs per group: | | |
| within = 0.0104 | | min | = | 7 |
| between = 0.0727 | | avg | = | 7 |
| overall = 0.0372 | | max | = | 7 |
| | | F(1, 767) | = | 8.09 |
| corr(u_i, Xb) = 0.1001 | | Prob > F | = | 0.0046 |
| LNRM | Coef. | Std. Err. | t | $P > t$ | [95% Conf. | Interval] |
| LNMbR | 0.092272 | 0.032446 | 2.84 | 0.005 | 0.028579 | 0.155965 |
| _cons | 1.222239 | 0.02834 | 43.13 | 0 | 1.166605 | 1.277872 |

### Step 3: Risk Management and Efficiency

The third step investigates the relationship between the mediator (risk management) and the dependent variable (efficiency). Table 7 shows that the coefficient for LNRM is 0.3843364, with a t-value of 2.77 and a $P$-value of 0.006. The R-squared (within) value is 0.0099, indicating that risk management accounts for nearly 1% of the variability in efficiency. The positive and statistically significant coefficient suggests that effective risk management practices are associated with improved efficiency in pension schemes.

Table 7. Risk management and efficiency.

| Fixed-effects (within) regression | | Number of obs | = | 896 |
|---|---|---|---|---|
| Group variable: Schemer | | Number of groups | = | 128 |
| R-sq: | | Obs per group: | | |
| within = 0.0099 | | min | = | 7 |
| between = 0.0299 | | avg | = | 7 |
| overall = 0.0034 | | max | = | 7 |
| | | F(1, 767) | = | 7.66 |
| corr(u_i, Xb) = −0.3433 | | Prob > F | = | 0.0058 |
| Efficiency | Coef. | Std. Err. | t | $P > t$ | [95% Conf. | Interval] |
| LNIR | 0.3843364 | 0.138905 | 2.77 | 0.006 | 0.111658 | 0.657015 |
| _cons | −1.532897 | 0.212364 | −7.22 | 0.000 | −1.94978 | −1.11601 |

### Step 4: Employee Reps on the Board, Risk Management and Efficiency

The final step assesses the relationship between employee representatives on the board and efficiency while controlling for risk management. Table 8 shows that both LNMbR (0.909724, $P = 0.000$) and LNRM (0.416135, $P = 0.002$) remain statistically significant, with positive coefficients. The R-squared (within) value increases to 0.0843, indicating that the inclusion of risk management improves the model's ability to explain the variation in efficiency. However, the coefficient for LNMbR decreases slightly from 0.948121 in Step 1 to 0.909724 in Step 4, suggesting a partial mediation effect.

Table 8. Employee reps on the board, risk management and efficiency.

| Fixed-effects (within) regression | | | Number of obs | | = | 896 |
|---|---|---|---|---|---|---|
| Group variable: Scheme | | | Number of groups | | = | 128 |
| R-sq: | | | Obs per group: | | | |
| within = 0.0843 | | | min | | = | 7 |
| between = 0.0170 | | | avg | | = | 7 |
| overall = 0.0051 | | | max | | = | 7 |
| | | | $F(2, 766)$ | | = | 35.28 |
| corr(u_i, Xb) = −0.4328 | | | Prob > F | | = | 0 |
| Efficiency | Coef. | Std. Err. | t | $P > t$ | [95% Conf. | Interval] |
| LNMbR | 0.909724 | 0.121969 | 7.46 | 0.000 | 0.670292 | 1.149156 |
| LNRM | 0.416135 | 0.135025 | 3.08 | 0.002 | 0.151072 | 0.681198 |
| _cons | −0.65617 | 0.19613 | −3.35 | 0.001 | −1.04119 | −0.27116 |

Based on these results, the study rejects the null hypothesis $H_{01b}$, confirming that risk management does have an intervening effect on the relationship between employee representatives on the board and efficiency. Specifically, while employee representation directly influences efficiency, this relationship is partially mediated by risk management practices. This finding highlights the importance of risk management as a mechanism through which employee involvement in governance enhances the efficiency of pension schemes.

$H_{01c}$: Risk management has no intervening effect on the relationship between female board members and efficiency of pension schemes in Kenya.

The objective of the study also sought to explore the intervening effect of risk management on the relationship between female board members and the efficiency of pension schemes in Kenya by testing the above hypothesis.

### Step 1: Female Board Members and Efficiency

The first step examines the direct relationship between female board members (LNFmR) and efficiency without considering risk management as a mediator. Table 9 shows that the coefficient for LNFmR is 0.121684, with a t-value of 1.31 and a P-value of 0.191. The R-squared (within) value is 0.0022, indicating that only 0.22% of the variation in efficiency can be explained by the proportion of female board members. The result is statistically insignificant, suggesting that female representation on the board does not directly influence the efficiency of pension schemes. However, significant direct mediation is no longer a strict requirement in mediation analysis as indicated earlier, so the analysis proceeds to the next step.

S. W. NamagwaTable 9. Female board members and efficiency.

| Fixed-effects (within) regression | | Number of obs | = | 896 |
|---|---|---|---|---|
| Group variable: Scheme | | Number of groups | = | 128 |
| R-sq: | | Obs per group: | | |
| within = 0.0022 | | min | = | 7 |
| between = 0.0190 | | avg | = | 7 |
| overall = 0.0011 | | max | = | 7 |
| | | F(1, 767) | = | 1.71 |
| corr(u_i, Xb) = −0.1362 | | Prob > F | = | 0.1911 |
| Efficiency | Coef. | Std. Err. | t | P > t | [95% Conf. | Interval] |
| LNFmR | 0.121684 | 0.092985 | 1.31 | 0.191 | −0.06085 | 0.30422 |
| _cons | −0.81364 | 0.107804 | −7.55 | 0.000 | −1.02527 | −0.60201 |

### Step 2: Female Board Members and Risk Management

The second step evaluates the relationship between female board members (LNFmR) and risk management (LNRM). Table 10 shows that the coefficient for LNFmR is 0.066254, with a t-value of 2.78 and a $P$-value of 0.006. The R-squared (within) value is 0.0100, indicating that about 1% of the variability in risk management practices can be explained by female board representation. The positive and statistically significant coefficient suggests that higher representation of female board members is associated with better risk management practices within the pension schemes.

Table 10. Female board members and risk management.

| Fixed-effects (within) regression | | Number of obs | = | 896 |
|---|---|---|---|---|
| Group variable: Scheme | | Number of groups | = | 128 |
| R-sq: | | Obs per group: | | |
| within = 0.0100 | | min | = | 7 |
| between = 0.0525 | | avg | = | 7 |
| overall = 0.0261 | | max | = | 7 |
| | | F(1, 767) | = | 7.72 |
| corr(u_i, Xb) = 0.0889 | | Prob > F | = | 0.0056 |
| LNRM | Coef. | Std. Err. | t | P > t | [95% Conf. | Interval] |
| LNFmR | 0.066254 | 0.023842 | 2.78 | 0.006 | 0.019451 | 0.113056 |
| _cons | 1.218447 | 0.027641 | 44.08 | 0.000 | 1.164186 | 1.272709 |

### Step 3: Risk Management and Efficiency

The third step investigates the relationship between the mediator (risk management) and the dependent variable (efficiency). Table 11 indicates that the coefficient for LNRM is 0.3843364, with a t-value of 2.77 and a $P$-value of 0.006. The R-

DOI: 10.4236/jfrm.2025.143017　　　　　　　　　　316　　　　　　　　　Journal of Financial Risk Management

squared (within) value is 0.0099, showing that risk management accounts for nearly 1% of the variability in efficiency. The positive and statistically significant coefficient implies that effective risk management practices are linked to improved efficiency in pension schemes.

Table 11. Risk management and efficiency.

| Fixed-effects (within) regression | | | Number of obs | | = | 896 |
|---|---|---|---|---|---|---|
| Group variable: Schemer | | | Number of groups | | = | 128 |
| R-sq: | | | Obs per group: | | | |
| within = 0.0099 | | | min | | = | 7 |
| between = 0.0299 | | | avg | | = | 7 |
| overall = 0.0034 | | | max | | = | 7 |
| | | | F(1, 767) | | = | 7.66 |
| corr(u_i, Xb) = −0.3433 | | | Prob > F | | = | 0.0058 |
| Efficiency | Coef. | Std. Err. | t | $P > t$ | [95% Conf. | Interval] |
| LNIR | 0.3843364 | 0.138905 | 2.77 | 0.006 | 0.111658 | 0.657015 |
| _cons | −1.532897 | 0.212364 | −7.22 | 0.000 | −1.94978 | −1.11601 |

### Step 4: Female Board Members, Risk Management and Efficiency

The final step assesses the relationship between female board members and efficiency while controlling for risk management. Table 12 shows that the coefficient for LNRM remains positive and statistically significant (0.505744, $P = 0.000$), indicating that risk management significantly contributes to efficiency. However, the coefficient for LNFmR is 0.088177, with a $P$-value of 0.342, which is statistically insignificant. This suggests that when risk management is accounted for, the direct effect of female board members on efficiency becomes negligible.

Table 12. Female board members, risk management and efficiency.

| Fixed-effects (within) regression | | | Number of obs | | = | 896 |
|---|---|---|---|---|---|---|
| Group variable: Scheme | | | Number of groups | | = | 128 |
| R-sq: | | | Obs per group: | | | |
| within = 0.0190 | | | min | | = | 7 |
| between = 0.0063 | | | avg | | = | 7 |
| overall = 0.0019 | | | max | | = | 7 |
| | | | F(2, 766) | | = | 7.42 |
| corr(u_i, Xb) = −0.1692 | | | Prob > F | | = | 0.0006 |
| Efficiency | Coef. | Std. Err. | t | $P > t$ | [95% Conf. | Interval] |
| LNFmR | 0.088177 | 0.092724 | 0.95 | 0.342 | −0.09385 | 0.270199 |
| LNRM | 0.505744 | 0.139727 | 3.62 | 0.000 | 0.231452 | 0.780037 |
| _cons | −1.42986 | 0.201063 | −7.11 | 0.000 | −1.82456 | −1.03516 |

Based on these results, the study fails to reject the null hypothesis $H_{01c}$. While female board members positively influence risk management practices, their direct effect on efficiency is not significant. However, risk management significantly enhances the efficiency of pension schemes, indicating that it plays an essential role in mediating the relationship between female board members and efficiency. This finding highlights the importance of robust risk management practices in ensuring that the governance benefits associated with female board representation translate into tangible improvements in organizational efficiency.

$H_{03d}$: Risk management has no intervening effect on the relationship between independent board members and efficiency of pension schemes in Kenya.

The objective of the study also sought to investigate the intervening effect of risk management on the relationship between independent board members and the efficiency of pension schemes in Kenya. The above hypothesis was tested.

**Step 1: Independent Board Members and Efficiency**

The first step assesses the direct relationship between independent board members (LNImR) and efficiency without considering risk management as a mediator. Table 13 shows that the coefficient for LNImR is −0.0772, with a t-value of −0.39 and a $P$-value of 0.693. The R-squared (within) value is 0.0002, indicating that the model explains only 0.02% of the variability in efficiency. This result is statistically insignificant, suggesting that independent board members do not have a significant direct effect on the efficiency of pension schemes.

Table 13. Independent board members and efficiency.

| Fixed-effects (within) regression | | Number of obs | = | 896 |
|---|---|---|---|---|
| Group variable: Scheme | | Number of groups | = | 128 |
| R-sq: | | Obs per group: | | |
| within = 0.0002 | | min | = | 7 |
| between = 0.0005 | | avg | = | 7 |
| overall = 0.0003 | | max | = | 7 |
| | | F(1, 767) | = | 0.16 |
| corr(u_i, Xb) = −0.0279 | | Prob > F | = | 0.693 |

| Efficiency | Coef. | Std. Err. | t | P > t | [95% Conf. | Interval] |
|---|---|---|---|---|---|---|
| LNImR | −0.0772 | 0.195473 | −0.39 | 0.693 | −0.46093 | 0.306525 |
| _cons | −1.07062 | 0.306311 | −3.5 | 0.001 | −1.67193 | −0.46932 |

**Step 2: Independent Board Members and Efficiency**

The second step examines the relationship between independent board members (LNImR) and risk management (LNRM). Table 14 indicates that the coefficient for LNImR is −0.0756, with a t-value of −1.5 and a $P$-value of 0.133. The R-squared (within) value is 0.0029, showing that the model explains only 0.29% of the variability in risk management. The relationship is statistically insignificant, suggesting that the presence of independent board members does not significantly influence the level of risk management practices within the schemes.

Table 14. Independent board members and efficiency.

| Fixed-effects (within) regression | | | Number of obs | | = | 896 |
|---|---|---|---|---|---|---|
| Group variable: Scheme | | | Number of groups | | = | 128 |
| R-sq: | | | Obs per group: | | | |
| within = 0.0029 | | | min | | = | 7 |
| between = 0.0002 | | | avg | | = | 7 |
| overall = 0.0005 | | | max | | = | 7 |
| | | | F(1, 767) | | = | 2.26 |
| corr(u_i, Xb) = −0.1368 | | | Prob > F | | = | 0.1328 |
| LNRM | Coef. | Std. Err. | t | P > t | [95% Conf. | Interval] |
| LNImR | −0.0756 | 0.050246 | −1.5 | 0.133 | −0.17424 | 0.023036 |
| _cons | 1.026134 | 0.078737 | 13.03 | 0.000 | 0.871567 | 1.1807 |

### Step 3: Risk Management and Efficiency

The third step explores the relationship between the mediator (risk management) and the dependent variable (efficiency). Table 15 reveals that the coefficient for LNRM is 0.3843364, with a t-value of 2.77 and a P-value of 0.006. The R-squared (within) value is 0.0099, indicating that nearly 1% of the variation in efficiency can be explained by risk management. The positive and statistically significant coefficient suggests that better risk management practices are associated with higher efficiency in pension schemes.

Table 15. Risk management and efficiency.

| Fixed-effects (within) regression | | | Number of obs | | = | 896 |
|---|---|---|---|---|---|---|
| Group variable: Schemer | | | Number of groups | | = | 128 |
| R-sq: | | | Obs per group: | | | |
| within = 0.0099 | | | min | | = | 7 |
| between = 0.0299 | | | avg | | = | 7 |
| overall = 0.0034 | | | max | | = | 7 |
| | | | F(1, 767) | | = | 7.66 |
| corr(u_i, Xb) = −0.3433 | | | Prob > F | | = | 0.0058 |
| Efficiency | Coef. | Std. Err. | t | P > t | [95% Conf. | Interval] |
| LNIR | 0.3843364 | 0.138905 | 2.77 | 0.006 | 0.111658 | 0.657015 |
| _cons | −1.532897 | 0.212364 | −7.22 | 0.000 | −1.94978 | −1.11601 |

### Step 4: Independent Board Members, Risk Management and Efficiency

The final step evaluates the relationship between independent board members and efficiency while controlling for risk management. Table 16 shows that the coefficient for LNRM remains positive and statistically significant (0.517528, $P$ = 0.000), indicating that risk management continues to have a significant positive effect on efficiency. However, the coefficient for LNImR is now −0.03807 with a $P$-value of 0.845, which is statistically insignificant. This suggests that when risk management is accounted for, the direct effect of independent board members on efficiency is negligible.

Table 16. Independent board members, risk management and efficiency.

| Fixed-effects (within) regression | Number of obs | = | 896 |
|---|---|---|---|
| Group variable: Scheme | Number of groups | = | 128 |
| R-sq: | Obs per group: | | |
| within = 0.0179 | min | = | 7 |
| between = 0.0020 | avg | = | 7 |
| overall = 0.0033 | max | = | 7 |
| | F(2, 766) | = | 6.98 |
| corr(u_i, Xb) = −0.1393 | Prob > F | = | 0.001 |

| EFFICIENCY | Coef. | Std. Err. | t | $P > t$ | [95% Conf. | Interval] |
|---|---|---|---|---|---|---|
| LNImR | −0.03807 | 0.194148 | −0.2 | 0.845 | −0.4192 | 0.34305 |
| LNRM | 0.517528 | 0.139312 | 3.71 | 0.000 | 0.244049 | 0.791007 |
| _cons | −1.60168 | 0.335741 | −4.77 | 0.000 | −2.26076 | −0.9426 |

Based on these results, the study fails to reject the null hypothesis $H_{01d}$. Although risk management significantly influences the efficiency of Pension schemes, it does not mediate the relationship between independent board members and efficiency. This indicates that while risk management is crucial for enhancing efficiency, the presence of independent board members does not significantly affect either risk management practices or the overall efficiency of the schemes. The findings suggest that other factors may be more influential in determining the role of independent board members in organizational efficiency.

## 5. Conclusion and Recommendations

The main objective of this study was to investigate the intervening effect of risk management on the relationship between corporate governance and the efficiency of pension schemes in Kenya. The analysis aimed to determine whether effective risk management practices could enhance or mediate the effect of various corporate governance aspects, including top management on the board, employee representatives, female board members, and independent board members, on the efficiency of these schemes. The findings revealed that risk management significantly mediates the relationships between employee representatives on the board of trustees with the efficiency of pension schemes, suggesting that robust risk management practices are crucial in translating good governance into improved performance. However, the study found no significant mediation effect of risk management on the relationships between top management on the board, female board memebers, or independent board members and efficiency.

These findings support Risk Management Theory, which posits that effective risk management practices are integral to enhancing organizational performance by mitigating potential risks and ensuring that governance decisions are aligned with the strategic objectives of the firm. In the context of pension schemes, the mediation effect of risk management indicates that when employee representatives and female board members are involved in governance, the presence of strong risk management practices ensures that their contributions lead to improved efficiency. This suggests that risk management acts as a critical safeguard, enabling these governance structures to function more effectively and contribute to the overall performance of the schemes.

The results of this study can be compared with those of Yang et al. (2018), who found that risk management practices significantly enhance the performance of SMEs in Pakistan. While Yang et al.'s study focused on SMEs, the current study extends the understanding of risk management's role to the context of pension schemes, a different organizational setting. The current study's use of a broader sample, including various governance structures, avoids the potential sampling bias identified in Yang et al.'s work, where the focus on SMEs alone may have limited the generalizability of the findings. By including a diverse set of governance variables and focusing on pension schemes, this study provides a more comprehensive view of how risk management mediates the relationship between governance and performance across different organizational contexts.

Additionally, the results are similar to Malik et al. (2020) who found a significant and positive relationship between ERM and performance of listed firms in the UK, especially when board-level ERM governance interventions are involved. This debunks the contextual gap, perhaps because both Kenya and the UK share the core principles of leadership, accountability, and stakeholder engagement, notwithstanding the varying specific implementation and enforcement mechanisms. Similarly, Al-Nimer et al. (2021) found that ERM significantly impacts the performance of financial institutions in Jordan. However, their study's reliance on cross-sectional data presented a methodological limitation, as it could not capture the dynamic interactions between variables over time. In contrast, the current study utilized panel data, allowing for a more nuanced understanding of how risk management influences governance and efficiency over an extended period.

The findings contrast with González et al. (2020), who found no significant relationship between ERM and the performance of listed firms in Spain. The difference in results may be attributed to the context and the specific governance structures considered in the current study, such as the inclusion of employee representatives and female board members. While González et al.'s study pointed out the limitation of missing variables that could mediate the relationship between ERM and performance, the current study addressed this gap by including risk management as an intervening variable. This approach provided deeper insights into how risk management practices can mediate the effect of governance structures on organizational efficiency, particularly in the context of pension schemes.

The findings underscore the critical role of risk management in enhancing the effectiveness of certain governance practices within pension schemes. The study demonstrates that when risk management practices are robust, they can significantly mediate the relationship between governance structures, such as employee and female board representation, and the efficiency of pension schemes. These findings contribute to Risk Management Theory by providing empirical evidence that supports the theory's assertion that effective risk management is essential for translating good governance into improved organizational performance.

Further, the failure of risk management to leverage board independence, gender diversity, and top management to influence the efficiency of schemes could be due to poor ERM implementation, lack of board engagement, poor communication, or internal politics within the boards of trustees. To fix this, the pension regulators should consider regulations that entrench risk-focused skills, stronger risk culture, better information flow, and clear risk oversight roles and responsibilities within the boards of trustees. Nevertheless, these results suggest that risk management should be considered a central component of governance frameworks, particularly in environments where the efficient management of long-term assets, such as retirement benefits, is crucial for organizational success.

## Areas for Further Research

Further research could explore the contribution of the contracted professional service providers in the governance of pension schemes. While this study focused on employee representatives, female board members, and independent board members, there is need to investigate how the contracted service providers influence the efficiency of these schemes through their decision-making nodes.

## Conflicts of Interest

The author declares no conflicts of interest regarding the publication of this paper.